\documentclass[pra,twocolumn,amsmath,18pt]{revtex4-1}
\usepackage{mathrsfs}
\usepackage{graphicx}
\usepackage{epstopdf}
\usepackage{amsthm}
\usepackage{amsmath}
\usepackage{amssymb}
\usepackage{braket}
\usepackage{bbold}
\usepackage{subfigure}
\usepackage{hyperref}
\usepackage{appendix}
\usepackage{tikz}

\DeclareMathOperator{\tr}{tr}

\def\be{\begin{equation}}
\def\ee{\end{equation}}
\def\ba{\begin{array}}
\def\ea{\end{array}}

\def\1{{\bf{1}}}

\input amssym.def
\begin{document}
\title{General monogamy and polygamy relations of arbitrary quantum
correlations for multipartite systems}

\smallskip
\author{Zhong-Xi Shen$^1$ }
\email{Corresponding author: 18738951378@163.com}
\author{Ke-Ke Wang$^1$}
\email{Corresponding author:  wangkk@cnu.edu.cn}
\author{Shao-Ming Fei$^{1,2}$}
\email{Corresponding author: feishm@cnu.edu.cn}
\affiliation{
$^1$School of Mathematical Sciences, Capital Normal University, Beijing 100048, China\\
$^2$Max-Planck-Institute for Mathematics in the Sciences, 04103 Leipzig, Germany
}

\begin{abstract}
Monogamy and polygamy of quantum correlations are the fundamental properties of
quantum systems. We study the monogamy and polygamy relations satisfied by any quantum correlations in multipartite quantum systems. General monogamy relations are presented for the $\alpha$th $(0\leq\alpha \leq\gamma$, $\gamma\geq2)$ power of quantum correlation, and general polygamy relations are given for the $\beta$th $(\beta\geq \delta$, $0\leq\delta\leq1)$ power of quantum correlation. We show that these newly derived monogamy and polygamy inequalities are tighter than the existing ones. By applying these results to specific quantum correlations such as concurrence and the square of convex-roof extended negativity of assistance (SCRENoA), the corresponding new classes of monogamy and polygamy relations are obtained, which include the existing ones as special cases. Detailed examples are given to illustrate the advantages of our results.\\

\end{abstract}

\maketitle

\section{Introduction}
Understanding the nature of quantum correlations is of significance in the study of quantum information processing. A fundamental property of quantum correlations is the monogamy for multipartite quantum systems. The monogamy of entanglement is a representative feature of quantum physics. It says that the entanglement of a quantum system with one of the other systems limits its entanglements with the remaining systems. The monogamy property is highly related to quantum information processing tasks such as the security analysis of quantum key distribution \cite{1}.

Determining whether or not a given quantum correlation measure is monogamous is an important question. Coffman, Kundu and Wootters(CKW) \cite{2} first characterized the monogamy of an entanglement measure $\mathcal{E}$ for three-qubit states $\rho_{ABC}$,
\begin{equation}
\mathcal{E}(\rho_{A|BC})\geq \mathcal{E}(\rho_{AB})+\mathcal{E}(\rho_{AC}),
\end{equation}
where $\rho_{AB}={\rm tr}_C(\rho_{ABC})$, $\rho_{AC}={\rm tr}_B(\rho_{ABC})$ are the reduced density matrices of $\rho_{ABC}$, $\mathcal{E}(\rho_{A|BC})$ stands for the entanglement under the bipartition $A$ and $BC$. It was showed that the squared concurrence \cite{3} and the squared convex-roof extended negativity (SCREN) \cite{4,5} satisfy the monogamy relations for multi-qubit states. Generalized families of tight monogamy inequalities related to the $\alpha$th ($\alpha\geq2$) power of concurrence \cite{6,7} and convex-roof extended negativity (CREN) \cite{7,8,9} have been then derived, which characterize finer entanglement distributions in multipartite systems.

The dual to the entanglement is the assisted entanglement, which may give rise to polygamy relations.  For arbitrary tripartite systems. Gour \emph{et.al} in Ref. \cite{10} established the first polygamy inequality by using the squared concurrence of assistance $C_a^2$. For a tripartite state $\rho_{ABC}$, the usual polygamy relation is characterized in the following form,
\begin{equation}
\mathcal{E}_{a}(\rho_{A|BC})\leq\mathcal{E}_{a} (\rho_{AB})+\mathcal{E}_{a}(\rho_{AC}),
\end{equation}
where $\mathcal{E}_{a}$ is the corresponding entanglement measure of assistance associated to $\mathcal{E}$. The authors in \cite{11} provided a characterization of multiqubit entanglement constraints in terms of the assistance of concurrence and CREN. By using SCRENoA and the Hamming weight of the binary vector related with the distribution of subsystems, the polygamy inequalities in terms of the $\beta$th power of SCRENoA for $0\leq\beta\leq1$ were established in \cite{12}. Similar to the monogamy relation, the corresponding tighter polygamy relations have also been established for different kinds of assisted entanglements \cite{13,14,15}.

Although many monogamy and polygamy relations have been proposed, the monogamy relations for the $\alpha$th $(0\leq\alpha\leq2)$ power and the polygamy relations for the $\beta$th $(\beta\geq1 )$ power of general quantum correlations in multipartite systems are still far from being satisfied \cite{16}. In this paper, we provide a class of monogamy and polygamy relations of the $\alpha$th $(0\leq\alpha\leq\gamma,\gamma\geq2)$ and the $\beta$th $(\beta\geq\delta,0\leq\delta\leq1)$ power for any quantum correlations. We also show that these new monogamy and polygamy inequalities are tighter than the existing ones. We illustrate these advantages with concrete examples for such as concurrence and the SCRENoA.

\section{\bf Preliminary knowledge}
Let $H_{A}$ and $H_{B}$ be $d_{A}$- and $d_{B}$-dimensional Hilbert spaces. The concurrence of a bipartite quantum pure state $|\varphi\rangle_{AB}\in H_{A}\otimes\ H_{B}$ is defined by \cite{17}
\begin{eqnarray}
C(|\varphi\rangle_{AB})=\sqrt{2[1-{\rm tr}(\rho_A^2)]},
\end{eqnarray}
where $\rho_{A}=\tr_{B}(\rho)$ is the reduced density matrix of $\rho=|\varphi\rangle_{AB}\langle\varphi|$. For a mixed bipartite quantum state $\rho_{AB}=\sum_ip_i|\varphi_i\rangle_{AB}\langle\varphi_i|\in H_{A}\otimes H_{B}$, the concurrence is given by the convex roof extension,
\begin{eqnarray}
C(\rho_{AB})=\min_{\{p_i,|\varphi_i\rangle\}}\sum_ip_iC(|\varphi_i\rangle_{AB}),
\end{eqnarray}
where the minimum is taken over all possible pure state decompositions of $\rho_{AB}=\sum_ip_i|\varphi_i\rangle_{AB}\langle\varphi_i|$.
For a 2-qubit mixed state $\rho$, the concurrence of $\rho$ is given by \cite{2},
\begin{eqnarray}
C(\rho)=\max\{\mu_1-\mu_2-\mu_3-\mu_4,0\},
\end{eqnarray}
where $\mu_1\geq\mu_2\geq\mu_3\geq\mu_4$ are the eigenvalues of the matrix $\sqrt{\sqrt{\rho}\widetilde{\rho}\sqrt{\rho}}$. Here, $\widetilde{\rho}=(\sigma_y\otimes\sigma_y) \rho^\ast (\sigma_y\otimes\sigma_y)$ and $\rho^\ast$ is the complex conjugation of $\rho$.

For an $N$-qubit quantum state $\rho_{AB_1B_2\cdots B_N-1}$, the concurrence under the bipartite partition $A$ and $B_1,B_2,\cdots,B_{N-1}$ satisfies the monogamy inequality for $\alpha\geq2$ \cite{6},
\begin{eqnarray}
C^\alpha(\rho_{A|B_1B_2\cdots B_N-1})&&\geq C^\alpha(\rho_{AB_{1}})
+C^\alpha(\rho_{AB_{2}})\nonumber\\ &&\quad +\cdots+C^\alpha(\rho_{AB_{N-1}}),
\end{eqnarray}
where $\rho_{AB_i}=\tr_{B_1\cdots B{i-1}B{i+1}\cdots B_{N-1}}(\rho_{AB_1 \cdots B_{N-1}})$, $i=1,\cdots,N-1$.

For a given state $\rho_{AB}\in H_A \otimes H_B$, the entanglement negativity is defined by \cite{18,19},
\begin{eqnarray}
\mathcal{N}(\rho_{AB})=\frac{\|\rho_{AB}^{T_{A}}\|_{\rm tr}-1}{2},
\end{eqnarray}
where $\rho_{AB}^{T_{A}}$ denotes the partial transpose of $\rho_{A}$ with respect to the subsystem $A$ and $\|\cdot\|_{\rm tr}$ denotes the trace norm. The trace norm is defined as the sum of the singular values of the matrix $M\in\mathbb{R}^{m\times n}$, i.e., $\|M\|_{\rm tr}=\sum_i\sigma_i={\rm tr}\sqrt{A^\dag A}$, where $\sigma_i$, $i=1,\cdots,\min(m,n)$, are the singular values of the matrix $A$ arranged in descending order. For convenience, we use $||\rho_{AB}^{T_A}||_{\rm tr}-1$ to define negativity in this paper.

For any bipartite quantum pure state $|\varphi\rangle_{AB}$ with Schmidt rank $d$, $|\varphi\rangle_{AB}=\sum_{i=1}^d\sqrt{\lambda_i}|ii\rangle$, its negativity is given by \cite{11},
\begin{eqnarray}
\mathcal{N}(|\varphi\rangle_{AB})=2\sum_{i<j}\sqrt{\lambda_i\lambda_j}=({\rm tr}\sqrt{\rho_{A}})^2-1,
\end{eqnarray}
where $\lambda_i, i=1,\cdots,d$, are the eigenvalues of the reduced density matrix $\rho_{A}={\rm tr}_{B}(|\varphi\rangle_{AB}\langle\varphi_i|)$.
For a mixed bipartite quantum state $\rho_{AB}=\sum_ip_i|\varphi_i\rangle\langle\varphi_i|\in H_{A}\otimes H_{B}$, the convex-roof extended negativity  is defined by \cite{20},
\begin{eqnarray}
\mathcal{N}(\rho_{AB})=\min_{\{p_i,|\varphi_i\rangle\}}\sum_ip_i\mathcal{N}(|\varphi_i\rangle),
\end{eqnarray}
where the minimum is taken over all possible convex partitions of $\rho_{AB}$ into pure state ensembles $\{p_i,|\varphi_i\rangle\}$, $0\leq p_i\leq1$ and $\sum_ip_i=1$. For any two-qubit mixed state $\rho_{AB}$, one has $\mathcal{N}(\rho_{AB})=C(\rho_{AB})$ \cite{13}. The convex-roof extended negativity of assistance (CRENoA) for mixed states is defined by \cite{12},
\begin{eqnarray}
\mathcal{N}_a(\rho_{AB})=\max_{\{p_i,|\varphi_i\rangle\}}\sum_ip_i\mathcal{N}(|\varphi_i\rangle),
\end{eqnarray}
where the maximum is taken over all possible convex partitions of $\rho_{AB}$ into pure state ensembles $\{p_i,|\varphi_i\rangle\}$, $0\leq p_i\leq1$ and $\sum_ip_i=1$. For any two-qubit mixed state $\rho_{AB}$, one has $\mathcal{N}_a(\rho_{AB})=C_a(\rho_{AB})$ \cite{4}.

For a mixed state $\rho_{AB}$, the square of convex-roof extended negativity (SCREN) is defined by
\begin{eqnarray}
\mathcal{N}_{sc}(\rho_{AB})=[\mathrm{min}\sum_ip_i\mathcal{N}(|\varphi_i\rangle_{AB})]^2.
\end{eqnarray}
%where the minimum is taken over all possible pure state decompositions $\{p_i,~|\varphi_i\rangle_{AB}\}$ of $\rho_{AB}$.
The square of convex-roof extended negativity of assistance  (SCRENoA) is then defined by
\begin{eqnarray}
\mathcal{N}_{sc}^a(\rho_{AB})=[\mathrm{max}\sum_ip_i\mathcal{N}(|\varphi_i\rangle_{AB})]^2.
\end{eqnarray}
%where the maximum is taken over all possible pure state decompositions $\{p_i,~|\varphi_i\rangle_{AB}\}$ of $\rho_{AB}$.
%For the $N$-qubit quantum state $\rho_{A|B_1\cdots B_{N-1}}$, regarded as a bipartite state under bipartite partition under the partition $A$ and $B_1,B_2,\cdots,B_{N-1}$,
The SCRENoA  satisfies the polygamy inequality for $0\leq\beta\leq1$ \cite{12},
\begin{eqnarray}
[\mathcal{N}_{sc}^{a}(\rho_{A|B_1\cdots B_{N-1}})]^{\beta}&\leq& \sum_{j=1}^{N-1}\beta^{j}[\mathcal{N}_{sc}^{a}(\rho_{AB_{j}})]^{\beta}\nonumber\\
&\leq& \sum_{j=1}^{N-1}[\mathcal{N}_{sc}^{a}(\rho_{AB_{j}})]^{\beta}.
\end{eqnarray}
%where $\rho_{AB_{j}}$ ($j=1,\cdots,N-1$) are the reduced density matrices of $\rho$, i.e, $\rho_{AB_j}=\Tr_{B_1\cdots B{j-1}B{j+1}\cdots B_{N-1}}(\rho_{AB_1 \cdots B_{N-1}})$.

Before giving our result, we propose the following lemma.

\noindent{[\bf Lemma 1]}.
Let $t$ ($t\geq1$) be real number. For any $x\geq t\geq1$, $1+\frac{1}{x}\leq q\leq 1+\frac{1}{t}$ and non-negative real numbers $m$ and $n$, we have
\begin{equation}\label{a1}
\begin{aligned}
(1+x)^{m}-q^{m-1}x^{m}\geq(1+t)^{m}-q^{m-1}t^{m}
\end{aligned}
\end{equation}
for $0\leq m\leq1$, and
\begin{equation}\label{a2}
\begin{aligned}
(1+x)^{n}-q^{n-1}x^{n}\leq(1+t)^{n}-q^{n-1}t^{n}
\end{aligned}
\end{equation}
for  $n\geq 1$.

\begin{proof}
We first consider the case for $0\leq m\leq1$. Consider the function  $f(x,m)=(1+x)^{m}-q^{m-1}x^{m}$ with $x\geq t\geq 1$, $1+\frac{1}{x}\leq p\leq 1+\frac{1}{t}$ and $0\leq m\leq1$. Since $\frac{\partial f(x,m)}{\partial x}=m(1+x)^{m-1}-mq^{m-1}x^{m-1}=m[(1+x)^{m-1}-(qx)^{m-1}]\geq0$, the function $f(x,m)$ is increasing with respect to $x$. As $x\geq t\geq 1$, $f(x,m)\geq f(t,m)=(1+t)^{m}-q^{m-1}t^{m}$, we get the inequality (\ref{a1}). Similar to the proof of the inequality (\ref{a1}), we can obtain the inequality (\ref{a2}).
\end{proof}

%For convenience, we denote by $C_{AB_i}=C(\rho_{AB_i})$ for $i=1,2,\cdots,N-1$,  $C_{A|B_1B_2\cdots B_{N-1}}=C(\rho_{A|B_1 B_2\cdots B_{N-1}})$, $N_{aAB_j}=\mathcal{N}_{sc}^{a}(\rho_{AB_j})$ for $j=1,2,\cdots,N-1$, and $N_{aA|B_1B_2\cdots B_{N-1}}=\mathcal{N}_{sc}^{a}(\rho_{A|B_1 B_2\cdots B_{N-1}})$ in the follwoing.

\section{MONOGAMY RELATIONS for GENERAL quantum correlations}
Let $\mathcal{Q}$ be any measure of quantum correlation for bipartite systems. $\mathcal{Q}$ is said to be monogamous if \cite{21},
\begin{eqnarray}\label{q1}
&&\mathcal{Q}(\rho_{A|B_1B_2,\cdots,B_{N-1}})\nonumber\\
&&\geq\mathcal{Q}(\rho_{AB_1})+\mathcal{Q}(\rho_{AB_2})+\cdots+\mathcal{Q}(\rho_{AB_{N-1}}).
\end{eqnarray}
For simplicity, we denote $\mathcal{Q}(\rho_{AB_i})$ by $\mathcal{Q}_{AB_i}$, and $\mathcal{Q}(\rho_{A|B_1B_2,\cdots,B_{N-1}})$ by $\mathcal{Q}_{A|B_1B_2,\cdots,B_{N-1}}$.
The relation (\ref{q1}) depends both on the measure $\mathcal{Q}$ and the state $\rho_{AB_1B_2,\cdots,B_{N-1}}$. Some of the quantum measures have been shown to be monogamous  for some classes states \cite{22,23}. However, there are other measures which does not satisfy the monogamy relations \cite{24,25}.

In Ref. \cite{26}, the authors proved that there exists $\gamma\in \mathbb{R}~(\gamma\geq2)$ such that for any $2\otimes2\otimes2^{N-2}$ tripartite state $\rho_{ABC} \in H_A\otimes H_B\otimes H_C$, $\mathcal{Q}$ satisfies the following monogamy relation,
\begin{eqnarray}\label{q2}
\mathcal{Q}^\gamma_{A|BC}\geq\mathcal{Q}^\gamma_{AB}+\mathcal{Q}^\gamma_{AC}.
\end{eqnarray}
The relation (\ref{q2}) holds for bipartite entanglement measures such as concurrence \cite{6}, entanglement of formation \cite{27} and convex-roof extended negativity \cite{27}.

Denote $\gamma$ the value such that $\mathcal{Q}$ satisfies the inequality (\ref{q2}). By using the inequality $(1+x)^m\geq 1+\frac{(1+k)^m-1}{k^m}x^m$ for $0\leq m \leq 1$, $x\geq k\geq1$,  the relation (\ref{q2}) is improved for $\gamma\geq2$ as
\begin{eqnarray}\label{q3}
\mathcal{Q}^\alpha_{A|BC}\geq\mathcal{Q}^\alpha_{AB}
+\frac{(1+k)^\frac{\alpha}{\gamma}-1}{k^\frac{\alpha}{\gamma}}\mathcal{Q}^\alpha_{AC}
\end{eqnarray}
with $\mathcal{Q}^\gamma_{AC}\geq k\mathcal{Q}^\gamma_{AB}$ and $0\leq\alpha\leq \gamma$ \cite{16}. By using the inequality $(1+x)^m\geq p^m+\frac{(1+k)^m-p^m}{k^m}x^m$ for $0\leq m \leq \frac{1}{2}$, $x\geq k\geq1$ and $\frac{1}{2}\leq p\leq1$, the relation (\ref{q3}) is improved for $\gamma\geq2$ as
\begin{eqnarray}\label{q4}
\begin{array}{rl}
&\mathcal{Q}_{A|BC}^\alpha \\ &\geq p^{\frac{\alpha}{\gamma}}\mathcal{Q}_{AB}^\alpha+\frac{(1+k)^{\frac{\alpha}{\gamma}}-p^{\frac{\alpha}{\gamma}}}{k^{\frac{\alpha}{\gamma}}}\mathcal{Q}_{AC}^\alpha
\end{array}
\end{eqnarray}
with $\mathcal{Q}^\gamma_{AC}\geq k\mathcal{Q}^\gamma_{AB}$ and  $0\leq\alpha\leq\frac{\gamma}{2}$ \cite{28}. By using the inequality $(1+x)^m\geq (1+a)^{m-1}+(1+\frac{1}{a})^{m-1}x^m$ for $0\leq m \leq 1$, $x\geq a\geq1$, the relation (\ref{q4}) is further improved for $\gamma\geq2$ as
\begin{equation}\label{q5}
\mathcal{Q}^{\alpha}_{A|BC}\geq (1+{a})^{\frac{\alpha}{\gamma}-1}\mathcal{Q}_{AB}^{\alpha}
+(1+\frac{1}{a})^{\frac{\alpha}{\gamma}-1}\mathcal{Q}_{AC}^{\alpha}
\end{equation}
with $\mathcal{Q}^\gamma_{AC}\geq a\mathcal{Q}^\gamma_{AB}$ and $0\leq\alpha\leq \gamma$ \cite{29}.

In the following, we propose a new class of monogamy relations of the $\alpha$th power of quantum correlation measure $\mathcal{Q}^\alpha$ for multipartite quantum states, which are tighter than the inequalities in Ref. \cite{29,16,28}.

\noindent{[\bf Theorem 1]}.
Let $\rho_{ABC}\in\mathcal{H}_A\otimes\mathcal{H}_B\otimes\mathcal{H}_C$ be a $2\otimes2\otimes2^{N-2}$ mixed state. If $\mathcal{Q}_{AC}^\gamma\geq t\mathcal{Q}_{AB}^\gamma$ and $1+\frac{\mathcal{Q}_{AB}^\gamma}{\mathcal{Q}_{AC}^\gamma}\leq q\leq1+\frac{1}{t}$ for $t\geq1$, we have
\begin{eqnarray}\label{q6}
\mathcal{Q}_{A|BC}^\alpha\geq ((1+t)^\frac{\alpha}{\gamma}
-q^{\frac{\alpha}{\gamma}-1}t^\frac{\alpha}{\gamma})
\mathcal{Q}_{AB}^\alpha+q^{\frac{\alpha}{\gamma}-1}\mathcal{Q}_{AC}^\alpha
\end{eqnarray}
for all $0\leq\alpha\leq\gamma$, $\gamma\geq2$.

\begin{proof}
By straightforward calculation, we have
\begin{eqnarray}\label{q7}
\mathcal{Q}_{A|BC}^\alpha&&=(\mathcal{Q}_{A|BC}^\gamma)^{\frac{\alpha}{\gamma}}
\geq(\mathcal{Q}_{AB}^\gamma+\mathcal{Q}_{AC}^\gamma)^{\frac{\alpha}{\gamma}}\nonumber\\
&&=\mathcal{Q}_{AB}^\alpha\Big(1+\frac{\mathcal{Q}_{AC}^\gamma}{\mathcal{Q}_{AB}^\gamma}\Big)^{\frac{\alpha}{\gamma}}\nonumber\\
&&\geq\mathcal{Q}_{AB}^\alpha\Big(q^{\frac{\alpha}{\gamma}-1}\frac{\mathcal{Q}_{AC}^\alpha}{\mathcal{Q}_{AB}^\alpha}
+(1+t)^\frac{\alpha}{\gamma}
-q^{\frac{\alpha}{\gamma}-1}t^\frac{\alpha}{\gamma}\Big)\nonumber\\
&&=((1+t)^\frac{\alpha}{\gamma}
-q^{\frac{\alpha}{\gamma}-1}t^\frac{\alpha}{\gamma})\mathcal{Q}_{AB}^\alpha+q^{\frac{\alpha}{\gamma}-1}\mathcal{Q}_{AC}^\alpha,\nonumber
\end{eqnarray}
where the second inequality is due to Lemma 1. Here, if $\mathcal{Q}_{AB}=0$, then $\mathcal{Q}_{AC}=0$, and the lower bound becomes trivially zero.
\end{proof}

\noindent{[\bf Remark 1]}. We have presented a universal form of monogamy relations that are complementary to the existing ones \cite{6,7,9} with different regions of the parameter $\alpha$ for any quantum correlations, which hold for any quantum correlation measures and any real number $\gamma$ satisfying the inequality (\ref{q2}). The new general monogamy relations apply to entanglement measures such as concurrence, entanglement of formation and the convex-roof extended negativity. These new monogamy relations can be also used for Tsallis-$q$ entanglement and Renyi-$q$ entanglement, and give rise to new monogamy relations.

\noindent{[\bf Remark 2]}. We have provided a class of lower bounds of the $\alpha$th power of quantum correlation measure $\mathcal{Q}$ with parameter $q$, where $1+\frac{\mathcal{Q}_{AB}^\gamma}{\mathcal{Q}_{AC}^\gamma}\leq q\leq1+\frac{1}{t}$. When $q=1+\frac{1}{t}$, $\mathcal{Q}_{AC}^\gamma\geq t\mathcal{Q}_{AB}^\gamma$ for all $0\leq\alpha\leq\gamma$ and $\gamma\geq2$, we have
\begin{eqnarray}\label{n1}
\mathcal{Q}^{\alpha}_{A|BC}\geq (1+{t})^{\frac{\alpha}{\gamma}-1}\mathcal{Q}_{AB}^{\alpha}
+(1+\frac{1}{t})^{\frac{\alpha}{\gamma}-1}\mathcal{Q}_{AC}^{\alpha}.
\end{eqnarray}
It is obvious that the relation (\ref{n1}) in \cite{29} is just a special case of our Theorem 1 when $t=a$. We see also that the lower bound of the $\alpha$th power of quantum correlation measure $\mathcal{Q}$ becomes tighter when $q$ decreases. Hence we get
\begin{eqnarray}
\mathcal{Q}_{A|BC}^\alpha&\geq& ((1+t)^\frac{\alpha}{\gamma}
-q^{\frac{\alpha}{\gamma}-1}t^\frac{\alpha}{\gamma})\mathcal{Q}_{AB}^\alpha
+q^{\frac{\alpha}{\gamma}-1}\mathcal{Q}_{AC}^\alpha\nonumber\\
&\geq& (1+{t})^{\frac{\alpha}{\gamma}-1}\mathcal{Q}_{AB}^{\alpha}
+(1+\frac{1}{t})^{\frac{\alpha}{\gamma}-1}\mathcal{Q}_{AC}^{\alpha}
\end{eqnarray}
for all $0\leq\alpha\leq\gamma$ and $\gamma\geq2$, where the the second equality holds when $q=1+\frac{1}{t}$. Namely, our inequality is tighter than that in \cite{29}. According to the results in Ref. \cite{29}, our monogamy relations are also better than the ones in \cite{16,28}.

Let us take concurrence as an example to illustrate the advantages of our new results. As the concurrence satisfies $C_{A|BC}^\gamma \geq C_{AB}^\gamma + C_{AC}^\gamma$ for any $2\otimes2\otimes2^{N-2}$ mixed tripartite state $\rho_{ABC}$ \cite{6} with $\gamma \geq 2$, we have the following corollary.

\noindent{[\bf Corollary 1]}.
For any $2\otimes2\otimes2^{N-2}$ mixed state $\rho_{ABC}\in\mathcal{H}_A\otimes\mathcal{H}_B\otimes\mathcal{H}_C$, if $C_{AC}^\gamma\geq tC_{AB}^\gamma$ and $1+\frac{C_{AB}^\gamma}{C_{AC}^\gamma}\leq q\leq1+\frac{1}{t}$ for $t\geq1$, we have
\begin{eqnarray}\label{q6}
C_{A|BC}^\alpha\geq ((1+t)^\frac{\alpha}{\gamma}
-q^{\frac{\alpha}{\gamma}-1}t^\frac{\alpha}{\gamma})C_{AB}^\alpha+q^{\frac{\alpha}{\gamma}-1}C_{AC}^\alpha
\end{eqnarray}
for all $0\leq\alpha\leq\gamma$, $\gamma\geq2$.\\

\noindent{[\bf Example 1]}. Consider the following three-qubit state $|\psi\rangle$ in generalized Schmidt decomposition \cite{30,31},
$$
|\psi\rangle=\lambda_{0}|000\rangle+\lambda_{1} e^{i \varphi}|100\rangle+\lambda_{2}|101\rangle+\lambda_{3}|110\rangle+\lambda_{4}|111\rangle,
$$
where $\lambda_{i} \geq 0$ and $\sum_{i=0}^{4} \lambda_{i}^{2}=1$. We have $C_{A|BC}=2 \lambda_{0} \sqrt{\lambda_{2}^{2}+\lambda_{3}^{2}+\lambda_{4}^{2}}$, $C_{AB}=2 \lambda_{0} \lambda_{2}$ and $C_{AC}=2 \lambda_{0} \lambda_{3}$.
Set $\lambda_{0}=\lambda_3=\frac{1}{2}$ and $\lambda_{1}=\lambda_{2}=\lambda_{4}=\frac{\sqrt{6}}{6}$.
We have
$C_{A|BC}=\frac{\sqrt{21}}{6}$, $C_{AB}=\frac{\sqrt{6}}{6}$ and $C_{AC}=\frac{1}{2}$. Setting $a=t=\sqrt{6}/2$, one gets the lower bound given in \cite{29},
\begin{eqnarray}\label{q8}
z_{1}&&=(1+{a})^{\frac{\alpha}{\gamma}-1}C_{AB}^{\alpha}
+(1+\frac{1}{a})^{\frac{\alpha}{\gamma}-1}C_{AC}^{\alpha}\nonumber\\
&&=(1+\frac{\sqrt{6}}{2})^{\frac{\alpha}{\gamma}-1}\left(\frac{\sqrt{6}}{6}\right)^{\alpha}\nonumber\\
&&\quad+(1+\frac{2}{\sqrt{6}})^{\frac{\alpha}{\gamma}-1}\left(\frac{1}{2}\right)^{\alpha}.
\end{eqnarray}
If we take $q=1+(\frac{\sqrt{6}}{3})^{\gamma}$, we get our lower bound
\begin{eqnarray}\label{q9}
z_{2}&&=((1+t)^\frac{\alpha}{\gamma}
-q^{\frac{\alpha}{\gamma}-1}t^\frac{\alpha}{\gamma})C_{AB}^\alpha+q^{\frac{\alpha}{\gamma}-1}C_{AC}^\alpha\nonumber\\
&&=\Big[(1+\frac{\sqrt{6}}{2})^{\frac{\alpha}{r}}-(1+(\frac{\sqrt{6}}{3})^{\gamma})^{\frac{\alpha}{r}-1}(\frac{\sqrt{6}}{2})^{\frac{\alpha}{r}}\Big]\left(\frac{\sqrt{6}}{6}\right)^{\alpha}\nonumber\\
&&\quad+(1+(\frac{\sqrt{6}}{3})^{\gamma})^{\frac{\alpha}{r}-1}\left(\frac{1}{2}\right)^{\alpha}.
\end{eqnarray}
It is seen that our bound is tighter than the one in \cite{29} for $0\leq\alpha\leq 2$ and $\gamma\geq 2$, see Fig 1. Fig. 2 shows the case of $\gamma=20$. Fig. 3 shows the difference $z=z_{2}-z_{1}$ of the concurrences between (\ref{q9}) and (\ref{q8}).
\begin{figure}[h]
\begin{center}
\includegraphics[width=8cm,clip]{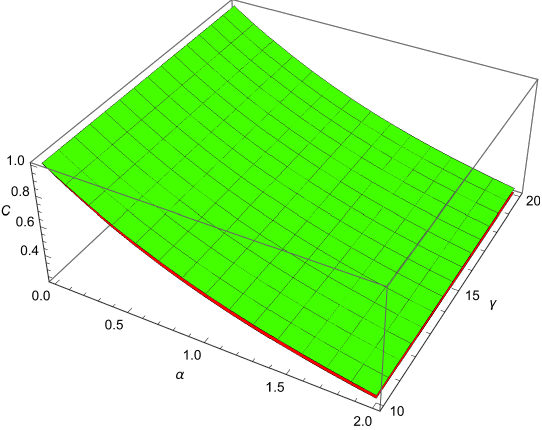}
\caption{The green surface represents
the lower bound from our result, red surface (below the
green one) represents the lower bound from the result in \cite{29}.}
\end{center}
\end{figure}
\begin{figure}[h]
\begin{center}
\includegraphics[width=8cm,clip]{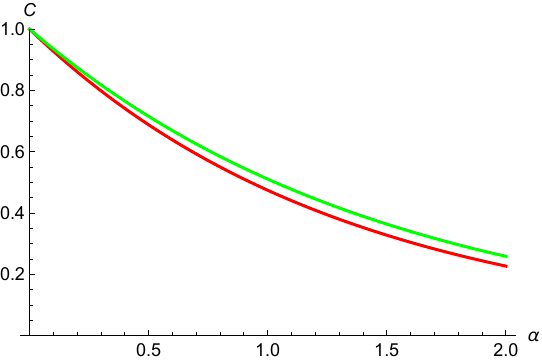}
\caption{The green (red) line represents
the lower bound from our result (\cite{29}).}
\end{center}
\end{figure}
\begin{figure}[h]
\begin{center}
\includegraphics[width=8cm,clip]{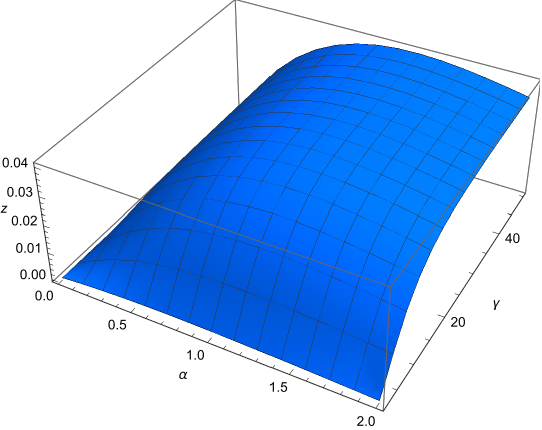}
\caption{The difference $z=z_{2}-z_{1}$, our lower bound  of concurrence minus that of \cite{29}.}
\end{center}
\end{figure}

Monogamy relations characterize the distributions of quantum correlations in multipartite systems and play a crucial role in the security of quantum cryptography.
Tighter monogamy relations imply finer characterizations of the quantum correlation distributions, which tighten the security bounds in quantum cryptography.
The complementary monogamy relations may also help to investigate the efficiency of entanglement used in quantum cryptography \cite{32} and in characterizations of the entanglement distributions.

Using the inequality $(1+t)^x\geq 1+t^x$ for $x\geq1,~0\leq t\leq1$, it is easy to generalize the result (\ref{q2}) to $N$-partite case,
\begin{eqnarray}\label{q10}
\mathcal{Q}^\gamma_{A|B_1B_2,\cdots,B_{N-1}}\geq \sum_{i=1}^{N-1}\mathcal{Q}_{AB_i}^\gamma.
\end{eqnarray}
By using Theorem 1 repeatedly, we have the following theorem for multipartite quantum systems.

\noindent{[\bf Theorem 2]}.
Let $t_{r}\geq1$ ($1\le r\le N-2$) be real number. For any $N$-qubit mixed state $\rho_{AB_1\cdots B_{N-1}}\in\mathcal{H}_A\otimes\mathcal{H}_{B_1}\otimes\cdots\otimes\mathcal{H}_{B_{N-1}}$, if $t_{i}\mathcal{Q}_{AB_i}^\gamma\leq \mathcal{Q}_{A|B_{i+1}\cdots B_{N-1}}^\gamma$,
$1+\frac{\mathcal{Q}_{AB_i}^\gamma}{\mathcal{Q}_{A|B_{i+1}\cdots B_{N-1}}^\gamma}\leq q_{i}\leq1+\frac{1}{t}$ for $i=1,2,\cdots,m$, and $t_{j}\mathcal{Q}_{A|B_{j+1}\cdots B_{N-1}}^{\gamma}\leq \mathcal{Q}_{AB_j}^\gamma$,
$1+\frac{\mathcal{Q}_{A|B_{j+1}\cdots B_{N-1}}^{\gamma}}{\mathcal{Q}_{AB_j}^\gamma}\leq q_{j}\leq1+\frac{1}{t}$ for $j=m+1,
\cdots,N-2$, $1\leq m\leq N-3,N\geq4$, we have
\begin{eqnarray}\label{m1}
&&\mathcal{Q}_{A|B_1\cdots B_{N-1}}^\alpha\nonumber\\
&&\geq l_{1}\mathcal{Q}_{AB_1}^\alpha+
\sum\limits_{i=2}^{m}\prod\limits_{h=1}^{i-1}q_{h}^{\frac{\alpha}{\gamma}-1}l_{i}\mathcal{Q}_{AB_{i}}^{\alpha}\nonumber\\
&&\quad+(q_1\cdots q_{m+1})^{\frac{\alpha}{\gamma}-1}
\mathcal{Q}_{AB_{m+1}}^{\alpha}\nonumber\\
&&\quad+(q_1\cdots q_{m})^{\frac{\alpha}{\gamma}-1}\Big(\sum\limits_{j=m+2}^{N-2}\prod\limits_{h=m+1}^{j-1}l_{h}q_{j}^{\frac{\alpha}{\gamma}-1}\mathcal{Q}_{AB_{j}}^{\alpha}\Big)\nonumber\\
&&\quad+(q_1\cdots q_{m})^{\frac{\alpha}{\gamma}-1}l_{m+1}\cdots l_{N-2}\mathcal{Q}_{AB_{N-1}}^\alpha
\end{eqnarray}
for all $0\leq\alpha\leq\gamma$, $\gamma\geq2$, where $l_{r}=(1+t_{r})^\frac{\alpha}{\gamma}
-q_{r}^{\frac{\alpha}{\gamma}-1}t_{r}^\frac{\alpha}{\gamma}$ with $1\le r\le N-2$.

\begin{proof}
From Theorem 1, if $t_{i}\mathcal{Q}_{AB_i}^\gamma\leq \mathcal{Q}_{A|B_{i+1}\cdots B_{N-1}}^\gamma$ and
$1+\frac{\mathcal{Q}_{AB_i}^\gamma}{\mathcal{Q}_{A|B_{i+1}\cdots B_{N-1}}^\gamma}\leq q_{i}\leq1+\frac{1}{t}$ for $i=1,2,\cdots,m$, we have
\begin{eqnarray}\label{m2}
&&\mathcal{Q}_{A|B_1\cdots B_{N-1}}^\alpha\nonumber\\
&&\geq l_{1}\mathcal{Q}_{AB_1}^\alpha+q_{1}^{\frac{\alpha}{\gamma}-1}\mathcal{Q}_{A|B_2\cdots B_{N-1}}^\alpha\nonumber\\
&&\geq l_{1}\mathcal{Q}_{AB_1}^\alpha+q_{1}^{\frac{\alpha}{\gamma}-1}l_{2}\mathcal{Q}_{AB_2}^\alpha
+(q_{1}q_{2})^{\frac{\alpha}{\gamma}-1}\mathcal{Q}_{A|B_3\cdots B_{N-1}}^\alpha\nonumber\\
&&\geq\cdots\nonumber\\
&&\geq l_{1}\mathcal{Q}_{AB_1}^\alpha+q_{1}^{\frac{\alpha}{\gamma}-1}l_{2}\mathcal{Q}_{AB_2}^\alpha+\cdots\nonumber\\
&&\quad+(q_{1}\cdots q_{m-1})^{\frac{\alpha}{\gamma}-1}l_{m}\mathcal{Q}_{AB_m}^\alpha\nonumber\\
&&\quad+(q_{1}\cdots q_{m-1})^{\frac{\alpha}{\gamma}-1}\mathcal{Q}_{A|B_{m+1}\cdots B_{N-1}}^\alpha.
\end{eqnarray}
If $t_{j}\mathcal{Q}_{A|B_{j+1}\cdots B_{N-1}}^{\gamma}\leq {Q}_{AB_j}^\gamma$ and
$1+\frac{\mathcal{Q}_{A|B_{j+1}\cdots B_{N-1}}^{\gamma}}{{Q}_{AB_j}^\gamma}\leq q_{j}\leq1+\frac{1}{t}$ for $j=m+1,
\cdots,N-2$, we get
\begin{eqnarray}\label{m3}
&&\mathcal{Q}_{A|B_{m+1}\cdots B_{N-1}}^\alpha\nonumber\\
&&\geq q_{m+1}^{\frac{\alpha}{\gamma}-1}\mathcal{Q}_{AB_{m+1}}^\alpha+l_{m+1}\mathcal{Q}_{A|B_{m+2}\cdots B_{N-1}}^\alpha\nonumber\\
&&\geq q_{m+1}^{\frac{\alpha}{\gamma}-1}\mathcal{Q}_{AB_{m+1}}^\alpha+l_{m+1}q_{m+2}^{\frac{\alpha}{\gamma}-1}\mathcal{Q}_{AB_{m+2}}^\alpha
\nonumber\\
&&\quad+l_{m+1}l_{m+2}\mathcal{Q}_{A|B_{m+3}\cdots B_{N-1}}^\alpha\geq\cdots\nonumber\\
&&\geq\cdots\nonumber\\
&&\geq q_{m+1}^{\frac{\alpha}{\gamma}-1}\mathcal{Q}_{AB_{m+1}}^\alpha+l_{m+1}q_{m+2}^{\frac{\alpha}{\gamma}-1}\mathcal{Q}_{AB_{m+2}}^\alpha+\cdots\nonumber\\
&&\quad+l_{m+1}\cdots l_{N-3}q_{N-2}^{\frac{\alpha}{\gamma}-1}\mathcal{Q}_{AB_{N-2}}^\alpha\nonumber\\
&&\quad+l_{m+1}\cdots l_{N-2}\mathcal{Q}_{AB_{N-1}}^\alpha.
\end{eqnarray}
Combining (\ref{m2}) and (\ref{m3}), we complete the proof.
\end{proof}

In particular, if $t_{i}\mathcal{Q}_{AB_i}^\gamma\leq \mathcal{Q}_{A|B_{i+1}\cdots B_{N-1}}^\gamma$ and
$1+\frac{\mathcal{Q}_{AB_i}^\gamma}{\mathcal{Q}_{A|B_{i+1}\cdots B_{N-1}}^\gamma}\leq q_{i}\leq1+\frac{1}{t}$ for $i=1,2,\cdots,N-2$, then the following monogamy relation holds.

\noindent{[\bf Theorem 3]}.
Let $t_{r}\geq1$ ($1\le r\le N-2$) be real number. For any $N$-qubit mixed state $\rho_{AB_1\cdots B_{N-1}}\in\mathcal{H}_A\otimes\mathcal{H}_{B_1}\otimes\cdots\otimes\mathcal{H}_{B_{N-1}}$, if $t_{i}\mathcal{Q}_{AB_i}^\gamma\leq \mathcal{Q}_{A|B_{i+1}\cdots B_{N-1}}^\gamma$ and
$1+\frac{\mathcal{Q}_{AB_i}^\gamma}{\mathcal{Q}_{A|B_{i+1}\cdots B_{N-1}}^\gamma}\leq q_{i}\leq1+\frac{1}{t}$ for $i=1,2,\cdots,N-2$, we have
\begin{eqnarray}\label{m1}
&&\mathcal{Q}_{A|B_1\cdots B_{N-1}}^\alpha\nonumber\\
&&\geq l_{1}\mathcal{Q}_{AB_1}^\alpha+
\sum\limits_{i=2}^{N-2}\prod\limits_{h=1}^{i-1}q_{h}^{\frac{\alpha}{\gamma}-1}l_{i}\mathcal{Q}_{AB_{i}}^{\alpha}\nonumber\\
&&\quad+(q_1\cdots q_{N-2})^{\frac{\alpha}{\gamma}-1}\mathcal{Q}_{AB_{N-1}}^\alpha,
\end{eqnarray}
for all $0\leq\alpha\leq\gamma$, $\gamma\geq2$, where $l_{r}=(1+t_{r})^\frac{\alpha}{\gamma}
-q_{r}^{\frac{\alpha}{\gamma}-1}t_{r}^\frac{\alpha}{\gamma}$.

For an $N$-qubit quantum state $\rho_{AB_1B_2\cdots B_{N-1}}$, it has been shown in \cite{6} that the $\alpha$th power of concurrence $C^\alpha$ ($0\leq\alpha\leq2$) does not satisfy the monogamy inequality $C^\alpha(|\psi\rangle_{AB_1B_2\cdots B_{N-1}})\geq\sum_{i=1}^{N-1}C^\alpha(\rho_{AB_i})$.
Set $\mathcal{Q}$ to be the concurrence of multipartite systems. Theorem 2 and Theorem 3 give new classes general monogamy inequality relations satisfied by the $\alpha$th power of concurrence for the case of $0\leq\alpha\leq\gamma$ and $\gamma\geq 2$. Similarly, the monogamy relations given in Theorem 2 and Theorem 3 can also be applied to other quantum correlations measures.

\section{polygamy RELATIONS for GENERAL quantum correlations}

The authors in \cite{33} proved that for arbitrary dimensional tripartite states, there exists $\delta\in \mathbb{R}~(0\leq\delta\leq1)$ such that a quantum correlation measure  $\mathcal{Q}$ satisfies the following polygamy relation,
\begin{eqnarray}\label{g1}
\mathcal{Q}^\delta_{A|BC}\leq\mathcal{Q}^\delta_{AB}+\mathcal{Q}^\delta_{AC}.
\end{eqnarray}
It has been shown that the assisted entanglement measures such as concurrence of assistance \cite{34}, entanglement of assistance \cite{35} and the convex-roof extended negativity of assistance \cite{33}, satisfy the polygamy relation (\ref{g1}).

In the following, we denote $\delta$ the value such that $\mathcal{Q}$ satisfies the inequality (\ref{g1}).  Using the inequality $(1+x)^m\leq 1+\frac{(1+k)^m-1}{k^m}x^m$ for $m \geq 1$, $x\geq k\geq1$,  the relation (\ref{g1}) is improved for $0\leq\delta\leq1$ as,
\begin{eqnarray}\label{g2}
\mathcal{Q}^\beta_{A|BC}\leq\mathcal{Q}^\beta_{AB}+\frac{(1+k)^\frac{\beta}{\delta}-1}{k^\frac{\beta}{\delta}}\mathcal{Q}^\beta_{AC}
\end{eqnarray}
with $\mathcal{Q}^\delta_{AC}\geq k\mathcal{Q}^\delta_{AB}$, and $\beta\geq\delta$ \cite{16}.
Using the inequality $(1+x)^m\leq p^m+\frac{(1+k)^m-p^m}{k^m}x^m$ for $m\geq1$, $x\geq k\geq1$ and $0\leq p\leq1$,  the relation (\ref{g2}) can be improved for $0\leq\delta\leq1$,
\begin{eqnarray}\label{g3}
\begin{array}{rl}
&\mathcal{Q}_{A|BC}^\beta \\ &\leq p^{\frac{\beta}{\delta}}\mathcal{Q}_{AB}^\beta+\frac{(1+k)^{\frac{\beta}{\delta}}-p^{\frac{\beta}{\delta}}}{k^{\frac{\beta}{\delta}}}\mathcal{Q}_{AC}^\beta
\end{array}
\end{eqnarray}
with $\mathcal{Q}^\delta_{AC}\geq k\mathcal{Q}^\delta_{AB}$ and $\beta\geq\delta$ \cite{28}.
Using the inequality $(1+x)^m\leq (1+a)^{m-1}+(1+\frac{1}{a})^{m-1}x^m$ for $m\geq1$, $x\geq a\geq1$,  the relation (\ref{g3}) is further improved for $0\leq\delta\leq1$,
\begin{equation}\label{g4}
\mathcal{Q}^{\beta}_{A|BC}\leq (1+{a})^{\frac{\beta}{\delta}-1}\mathcal{Q}_{AB}^{\beta}+(1+\frac{1}{a})^{\frac{\beta}{\delta}-1}\mathcal{Q}_{AC}^{\beta}
\end{equation}
with $\mathcal{Q}^\delta_{AC}\geq a\mathcal{Q}^\delta_{AB}$ and $\beta\geq\delta$ \cite{29}.

In the following, for the $\beta$th power of  quantum correlation measure $\mathcal{Q}^\beta$, we propose a new class of polygamy relations for multipartite quantum states according to Lemma 1, which are tighter than the inequalities in Ref. \cite{29,16,28}.
Using the similar approach in the proof of Theorem 1 and the Lemma 1, we have

\noindent{[\bf Theorem 4]}.
For any $2\otimes2\otimes2^{N-2}$ mixed state $\rho_{ABC}\in\mathcal{H}_A\otimes\mathcal{H}_B\otimes\mathcal{H}_C$, if $\mathcal{Q}_{AC}^\delta\geq t\mathcal{Q}_{AB}^\delta$ and $1+\frac{\mathcal{Q}_{AB}^\delta}{\mathcal{Q}_{AC}^\delta}\leq p\leq1+\frac{1}{t}$ for $t\geq1$, we have
\begin{eqnarray}\label{g5}
\mathcal{Q}_{A|BC}^\beta\leq ((1+t)^\frac{\beta}{\delta}
-q^{\frac{\beta}{\delta}-1}t^\frac{\beta}{\delta})\mathcal{Q}_{AB}^\beta
+q^{\frac{\beta}{\delta}-1}\mathcal{Q}_{AC}^\beta
\end{eqnarray}
for all $\beta\geq\delta$ and $0\leq\delta\leq1$.

\noindent{[\bf Remark 3]}.
We have presented a class of upper bounds of the $\beta$th power of any quantum correlation measure $\mathcal{Q}$ with parameter $q$ satisfying $1+\frac{\mathcal{Q}_{AB}^\delta}{\mathcal{Q}_{AC}^\delta}\leq q\leq1+\frac{1}{t}$. When $q=1+\frac{1}{t}$, $\mathcal{Q}_{AC}^\delta\geq t\mathcal{Q}_{AB}^\delta$ for all $\beta\geq\delta$ and $0\leq\delta\leq1$, we have
\begin{eqnarray}\label{n2}
\mathcal{Q}^{\beta}_{A|BC}\leq (1+{t})^{\frac{\beta}{\delta}-1}\mathcal{Q}_{AB}^{\beta}+(1+\frac{1}{t})^{\frac{\beta}{\delta}-1}\mathcal{Q}_{AC}^{\beta}.
\end{eqnarray}
If $t=a$, one sees that the relation (\ref{n2}) in \cite{29} is just a special case of our Theorem 4. Moreover, it is seen that the upper bound of the $\beta$th power of quantum correlation measure $\mathcal{Q}$ gets tighter when $q$ increases. Hence, we obtain
\begin{eqnarray}
\mathcal{Q}_{A|BC}^\beta&\leq& ((1+t)^\frac{\beta}{\delta}
-q^{\frac{\beta}{\delta}-1}t^\frac{\beta}{\delta})\mathcal{Q}_{AB}^\beta+q^{\frac{\beta}{\delta}-1}\mathcal{Q}_{AC}^\beta\nonumber\\
&\leq& (1+{t})^{\frac{\beta}{\delta}-1}\mathcal{Q}_{AB}^{\beta}
+(1+\frac{1}{t})^{\frac{\beta}{\delta}-1}\mathcal{Q}_{AC}^{\beta}
\end{eqnarray}
for all $\beta\geq\delta$ and $0\leq\delta\leq1$, where the second equality holds when $q=1+\frac{1}{t}$. Therefore, our inequality is tighter than that in \cite{29}, and hence also than that in \cite{16,28}.

We have presented a general polygamy relation for any quantum correlation measure and real number $\delta$ satisfying inequality (\ref{g1}). The new general polygamy relation can be applied to assisted entanglement measures like concurrence of assistance, entanglement of assistance and the square of convex-roof extended negativity of assistance (SCRENoA),as well as  the Tsallis-$q$ entanglement of assistance and Renyi-$q$ entanglement of assistance, which can  give new polygamy relations. Let us take the SCRENoA as an example to illustrate the advantages of our new results. Since SCRENoA satisfies $N_{aA|BC}^\delta \leq N_{aAB}^\delta + N_{aAC}^\delta$ for any $2\otimes2\otimes2^{N-2}$ mixed tripartite state $\rho_{ABC}$ \cite{12} with $0\leq\delta\leq1$, we have

\noindent{[\bf Corollary 4]}.
For any $2\otimes2\otimes2^{N-2}$ mixed state $\rho_{ABC}\in\mathcal{H}_A\otimes\mathcal{H}_B\otimes\mathcal{H}_C$, if $N_{aAC}^\delta\geq tN_{aAB}^\delta$ and $1+\frac{N_{aAB}^\delta}{N_{aAC}^\delta}\leq p\leq1+\frac{1}{t}$ for $t\geq1$, then
\begin{eqnarray}\label{g6}
N_{aA|BC}^\beta\leq ((1+t)^\frac{\beta}{\delta}
-q^{\frac{\beta}{\delta}-1}t^\frac{\beta}{\delta})N_{aAB}^\beta
+q^{\frac{\beta}{\delta}-1}N_{aAC}^\beta
\end{eqnarray}
for all $\beta\geq\delta$, $0\leq\delta\leq1$.

\noindent{[\bf Example 2]}.
Let us consider the three-qubit generlized $W$-class state,
\begin{eqnarray}\label{}
|W\rangle_{ABC}=\frac{1}{2}(|100\rangle+|010\rangle)+\frac{\sqrt{2}}{2}|001\rangle.
\end{eqnarray}
Then $N_{aABC}=\frac{3}{4}$, $N_{aAB}=\frac{1}{4}$ and $N_{aAC}=\frac{1}{2}$.
Set $a=t=2^{0.6}$. The upper bound given in \cite{29} is
\begin{eqnarray}\label{g7}
W_{1}&&=(1+{a})^{\frac{\beta}{\delta}-1}N_{aAB}^{\beta}+(1+\frac{1}{a})^{\frac{\beta}{\delta}-1}N_{aAC}^{\beta}\nonumber\\
&&=(1+2^{0.6})^{\frac{\beta}{\delta}-1}(\frac{1}{4})^\beta\nonumber\\
&&\quad+(1+\frac{1}{2^{0.6}})^{\frac{\beta}{\delta}-1} (\frac{1}{2})^\beta.
\end{eqnarray}
Take $q=1+(\frac{1}{2})^{\delta}$. Our upper bound is
\begin{eqnarray}\label{g8}
W_{2}&&=((1+t)^\frac{\beta}{\delta}
-q^{\frac{\beta}{\delta}-1}t^\frac{\beta}{\delta})N_{aAB}^\beta+q^{\frac{\beta}{\delta}-1}N_{aAC}^\beta\nonumber\\
&&=\Big[(1+2^{0.6})^{\frac{\beta}{\delta}}-(1+(\frac{1}{2})^{\delta})^{\frac{\beta}{\delta}-1}(2^{0.6})^{\frac{\beta}{\delta}}\Big]
\left(\frac{\sqrt{6}}{6}\right)^{\beta}\nonumber\\
&&\quad+(1+(\frac{1}{2})^{\delta})^{\frac{\beta}{\delta}-1}\left(\frac{1}{2}\right)^{\beta}.
\end{eqnarray}
It can be verified that our bound is better
than the one in \cite{29} for $\beta\geq\delta$ and $0\leq\delta\leq1$, see Fig. 4.
Fig. 5 shows the case of $\delta=0.8$, and Fig. 6 shows the difference $W=W_{1}-W_{2}$ of the SCRENoA between (\ref{g7}) and (\ref{g8}).
\begin{figure}[h]
\begin{center}
\includegraphics[width=8cm,clip]{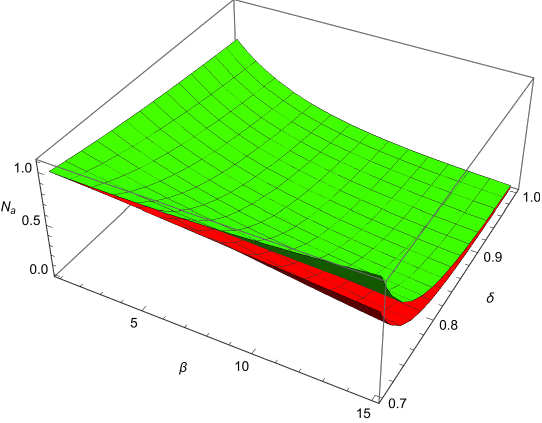}
\caption{The red surface represents
the upper bound from our result, green surface  represents the upper bound from the result in \cite{29}.}
\end{center}
\end{figure}
\begin{figure}[h]
\begin{center}
\includegraphics[width=8cm,clip]{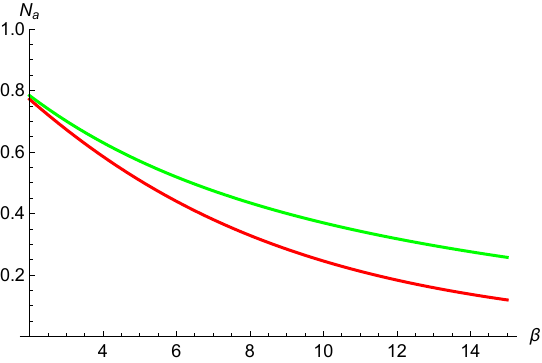}
\caption{The red line represents the upper bound from our result, the green line represents the upper bound from the result in \cite{29}.}
\end{center}
\end{figure}
\begin{figure}[h]
\begin{center}
\includegraphics[width=8cm,clip]{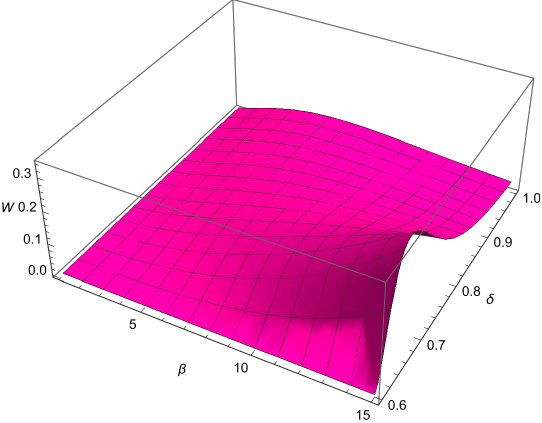}
\caption{The surface stands for the upper bound of SCRENoA \cite{29} minus our upper bound.}
\end{center}
\end{figure}

Using the inequality $(1+t)^x\leq 1+t^x$ for $0\leq x\leq1,~0\leq t\leq1$, it is easy to generalize the result (\ref{g1}) to $N$-partite case,
\begin{eqnarray}
\mathcal{Q}^\delta_{A|B_1B_2,\cdots,B_{N-1}}\leq \sum_{i=1}^{N-1}\mathcal{Q}_{AB_i}^\delta.
\end{eqnarray}
By using Theorem 4 repeatedly, with the similar method to the proof of Theorem 2, we have the following theorem for multipartite quantum systems.

\noindent{[\bf Theorem 5]}.
Let $t_{r}\geq1$ ($1\le r\le N-2$) be real number. For any $N$-qubit mixed state $\rho_{AB_1\cdots B_{N-1}}\in\mathcal{H}_A\otimes\mathcal{H}_{B_1}\otimes\cdots\otimes\mathcal{H}_{B_{N-1}}$, if $t_{i}\mathcal{Q}_{AB_i}^\delta\leq \mathcal{Q}_{A|B_{i+1}\cdots B_{N-1}}^\delta$,
$1+\frac{\mathcal{Q}_{AB_i}^\delta}{\mathcal{Q}_{A|B_{i+1}\cdots B_{N-1}}^\delta}\leq q_{i}\leq1+\frac{1}{t}$ for $i=1,2,\cdots,m$, and $t_{j}\mathcal{Q}_{A|B_{j+1}\cdots B_{N-1}}^{\delta}\leq \mathcal{Q}_{AB_j}^\delta$,
$1+\frac{\mathcal{Q}_{A|B_{j+1}\cdots B_{N-1}}^{\delta}}{\mathcal{Q}_{AB_j}^\delta}\leq q_{j}\leq1+\frac{1}{t}$ for $j=m+1,
\cdots,N-2$, $1\leq m\leq N-3,N\geq4$, we have
\begin{eqnarray}\label{g9}
&&\mathcal{Q}_{A|B_1\cdots B_{N-1}}^\beta\nonumber\\
&&\leq \mathcal{K}_{1}\mathcal{Q}_{AB_1}^\beta+
\sum\limits_{i=2}^{m}\prod\limits_{h=1}^{i-1}q_{h}^{\frac{\beta}{\delta}-1}\mathcal{K}_{i}\mathcal{Q}_{AB_{i}}^{\beta}\nonumber\\
&&\quad+(q_1\cdots q_{m+1})^{\frac{\beta}{\delta}-1}
\mathcal{Q}_{AB_{m+1}}^{\beta}\nonumber\\
&&\quad+(q_1\cdots q_{m})^{\frac{\beta}{\delta}-1}\Big(\sum\limits_{j=m+2}^{N-2}\prod\limits_{h=m+1}^{j-1}\mathcal{K}_{h}q_{j}^{\frac{\beta}{\delta}-1}\mathcal{Q}_{AB_{j}}^{\beta}\Big)\nonumber\\
&&\quad+(q_1\cdots q_{m})^{\frac{\beta}{\delta}-1}\mathcal{K}_{m+1}\cdots \mathcal{K}_{N-2}\mathcal{Q}_{AB_{N-1}}^\beta
\end{eqnarray}
for all $\beta\geq\delta$, $0\leq\delta\leq1$, where $\mathcal{K}_{r}=(1+t_{r})^\frac{\beta}{\delta}
-q_{r}^{\frac{\beta}{\delta}-1}t_{r}^\frac{\beta}{\delta}$ with $1\le r\le N-2$.

In particular, we have the following polygamy relation.

\noindent{[\bf Theorem 6]}.
Let $t_{r}\geq1$ ($1\le r\le N-2$) be real number. For any $N$-qubit mixed state $\rho_{AB_1\cdots B_{N-1}}\in\mathcal{H}_A\otimes\mathcal{H}_{B_1}\otimes\cdots\otimes\mathcal{H}_{B_{N-1}}$, if $t_{i}\mathcal{Q}_{AB_i}^\delta\leq \mathcal{Q}_{A|B_{i+1}\cdots B_{N-1}}^\delta$ and
$1+\frac{\mathcal{Q}_{AB_i}^\delta}{\mathcal{Q}_{A|B_{i+1}\cdots B_{N-1}}^\delta}\leq q_{i}\leq1+\frac{1}{t}$ for $i=1,2,\cdots,N-2$, we have
\begin{eqnarray}\label{g10}
&&\mathcal{Q}_{A|B_1\cdots B_{N-1}}^\beta\nonumber\\
&&\leq \mathcal{K}_{1}\mathcal{Q}_{AB_1}^\beta+
\sum\limits_{i=2}^{N-2}\prod\limits_{h=1}^{i-1}q_{h}^{\frac{\beta}{\delta}-1}\mathcal{K}_{i}\mathcal{Q}_{AB_{i}}^{\beta}\nonumber\\
&&\quad+(q_1\cdots q_{N-2})^{\frac{\beta}{\delta}-1}\mathcal{Q}_{AB_{N-1}}^\beta
\end{eqnarray}
for all $\beta\geq\delta$, $0\leq\delta\leq1$, where $\mathcal{K}_{r}=(1+t_{r})^\frac{\beta}{\delta}
-q_{r}^{\frac{\beta}{\delta}-1}t_{r}^\frac{\beta}{\delta}$ with $1\le r\le N-2$.

We use again the SCRENoA as an example to demonstrate the advantages of our polygamy relations for multipartite systems. For an $N$-qubit quantum state $\rho_{AB_1B_2\cdots B_{N-1}}$, it has been shown in \cite{12} that the $\beta$th power of SCRENoA $N_{a}^{\beta}$ ($\beta\geq1$) does not satisfy the polygamy inequalities $N_{aAB_1B_2\cdots B_{N-1}}^{\beta}\leq\sum_{i=1}^{N-1}N_{aAB_{i}}^{\beta}$. Our Theorem 5 and 6 give new classes general polygamy inequalities satisfied by the $\beta$th power of SCRENoA for the case of $\beta\geq\delta$, $0\leq\delta\leq1$.

\section{conclusion}
Monogamy and polygamy inequalities of quantum correlations characterize the fundamental properties of quantum systems. We have investigated the monogamy and polygamy relations satisfied by any quantum correlation measures in arbitrary dimensional multipartite quantum systems. General monogamy relations are obtained for the $\alpha$th $(0\leq\alpha\leq\gamma$, $\gamma\geq2)$ power of quantum correlations, as well as for the polygamy relations of the $\beta$th $(\beta \geq\delta$, $0\leq\delta\leq 1)$ power. These monogamy and polygamy inequalities are complementary to the existing ones with different regions of $\alpha$ and $\beta$. Applying the general quantum correlations to specific quantum correlations, the corresponding new class of monogamy and polygamy relations are obtained, which include the existing ones as special cases. We have also used the concurrence and the SCRENoA to show that our bounds are indeed better than the recently available bounds by detailed examples. Our results may provide a rich reference for future works on the study of multiparty quantum correlations.

\bigskip
\noindent{\bf Acknowledgments}
This work is supported by the National Natural Science Foundation of China (NSFC) under Grants 12075159 and 12171044, Beijing Natural Science Foundation (Grant No. Z190005), the Academician Innovation Platform of Hainan Province.

\bigskip
\noindent{\bf Data availability statement}
All data generated or analysed during this study are included in this published article.

%\bigskip
%\noindent{\bf  Declarations}

%\textbf{Conflict of interest}
%The authors declare no competing interests.

\end{document}